\documentclass[a4paper]{jpconf}

\newcommand\Rey{\mbox{Re}} 
\usepackage{amsmath,amssymb}
\usepackage{graphicx}
\usepackage{natbib}
\begin{document}

\title{Modeling the transition to turbulence in shear flows}
\author{Dwight Barkley$^{1,2}$}

\address{$^1$ Mathematics Institute, University of Warwick, 
Coventry CV4 7AL, United Kingdom}
\address{$^2$ PMMH (UMR 7636 CNRS - ESPCI - Univ Paris 06 - Univ Paris 07), 
10 rue Vauquelin, 75005 Paris, France}

\ead{D.Barkley@warwick.ac.uk}

\begin{abstract}
One-dimensional models are presented for transitional shear flows.  The models
have two variables corresponding to turbulence intensity and mean shear. These
variables evolve according to simple equations based on known properties of
transitional turbulence. The first model considered is for pipe flow. A
previous study modeled turbulence using a chaotic tent map. In the present
work turbulence is modeled instead as multiplicative noise. This model
captures the character of transitional pipe flow and contains metastable
puffs, puff splitting, and slugs. These ideas are extended to a limited model
of plane Couette flow.
\end{abstract}

\section{Introduction}

The transition to turbulence in shear flows has been the subject of study
since Reynolds' pioneering studies over a century ago~\citep{Reynolds:1883}.
The difficulty in understanding shear-flow transition is largely attributable
to the subcritical nature of the problem.  In all of the classic cases -- pipe
flow, channel flow, plane Couette flow, boundary layer flow and others --
turbulence is found at Reynolds numbers for which laminar flow is linearly
stable.  In such flows turbulence appears abruptly following finite-sized
disturbances of laminar flow and not through a sequence of instabilities each
increasing the dynamical complexity of the flow.  This limits the
applicability of linear and weakly nonlinear theories in addressing transition
in these cases.

One of the most intriguing aspects of shear turbulence is the intermittent form
it takes in the transitional regime, near the minimum Reynolds number for
which turbulence can be triggered.  In pipe flow, one observes localized
turbulent patches, known as puffs, surrounded upstream and downstream by
laminar flow~\citep{Wygnanski:1973, Nishi:2008, Mullin:2011}.  In planar
cases, such as plane Couette flow and boundary layer flow, one commonly
observes turbulent spots surrounded by laminar flow
\citep{Wygnanski:1976,Tillmark:1992}. Even more intriguing is
the regular alternation of turbulent and laminar regions that is now known to
arise spontaneously in many shear flows with sufficiently large aspect
ratio~\citep{Prigent:2002,Barkley:2005p1065}.

Minimal models of spatiotemporal intermittency have been useful in
understanding generic features of intermittent shear
turbulence~\citep{Chate:1988p8,Bottin:1998p305}.  Here, I consider models that
contain more of the physics specific to shear turbulence and from this I
obtain models that produce quite realistic dynamics. For pipe flow it is
possible to reproduce nearly all of the large-scale phenomena associated with
transition using only two scalar equations.
Other shear flows are more difficult, but I point to some ideas for plane
Couette flow.

\section{Pipe flow}

In this section I consider pipe flow. As concerns the large-scale features,
pipe flow is effectively a one-dimensional system and this makes it a
particularly good problem to tackle first. I will summarize basic features of
pipe flow and recall the modeling proposed in \citet{barkley:2011}. Then I
will consider an alternative approach to that in \citet{barkley:2011} and here
model turbulence by multiplicative noise.

\subsection{Phenomenology}

\begin{figure}[t]
\begin{minipage}{14pc}
\includegraphics[width=14pc]{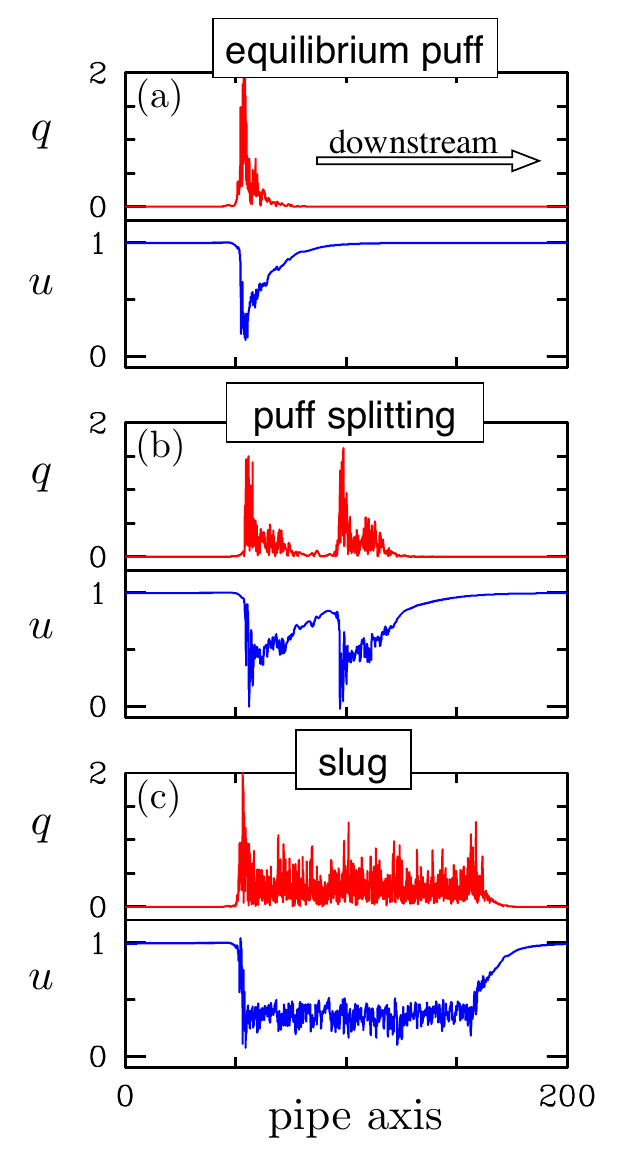}
\caption{\label{fig:DNS}
Regimes of transitional pipe flow from simulations in a periodic pipe 200D
long. Shown are instantaneous values of turbulence intensity $q$ and axial
velocity $u$ along the pipe axis.  (a) Equilibrium puff at $\Rey = 2000$.  
(b) Puff splitting at $\Rey = 2275$. 
(c) Slug flow at $\Rey = 3200$.}
\end{minipage}\hspace{2.9pc}%
\begin{minipage}{21pc}
\includegraphics[width=21pc]{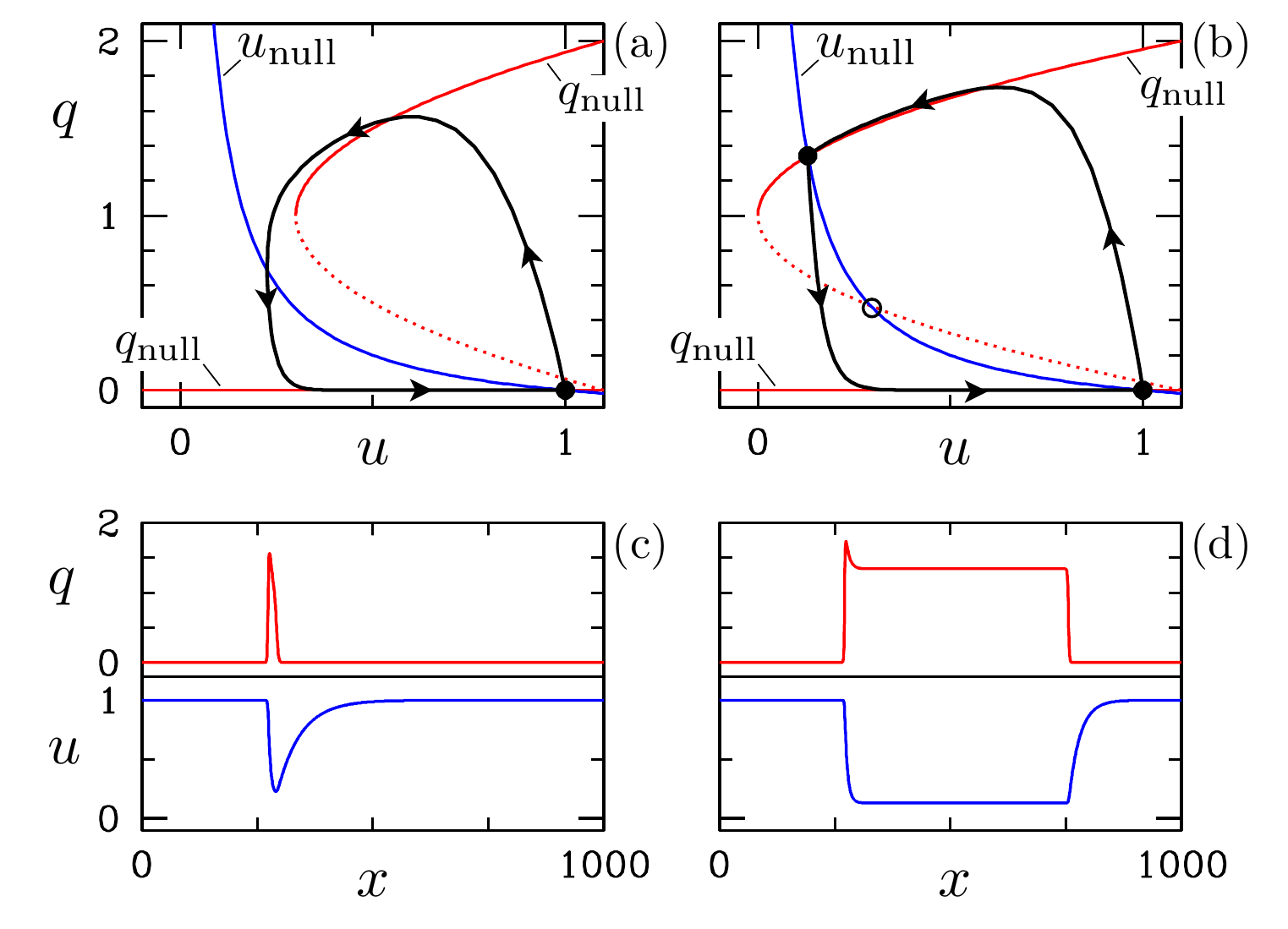}
\caption{\label{fig:pde}
The distinction between puffs and slugs seen as the difference between
excitability and bistablilty in the PDE model,
Eqs.~\eqref{eq:q_pde}-\eqref{eq:u_pde}. Phase planes show nullclines at (a) $r
= 0.7$ and (b) $r = 1$. The fixed point $(1,0)$ corresponds to stable laminar
flow. In (a) this is the only fixed point. In (b) the additional stable fixed
point corresponds to stable turbulence. Solution snapshots show (c) a puff at
$r = 0.7$ and (d) a slug at $r = 1$. These solutions are plotted in the phase
planes with arrows indicating increasing $x$.  }
\end{minipage} 
\end{figure}

Figure~\ref{fig:DNS} summarizes the three important dynamical regimes of
transitional pipe flow from direct numerical simulations
(DNS)~\citep{Blackburn:2004p1087,Moxey:2010}.  Quantities are
nondimensionalized by the pipe diameter $D$ and the mean (bulk) velocity $\bar
U$.  The Reynolds numbers is $\Rey = D\bar U/\nu$, where $\nu$ is kinematic
viscosity.  Flows are well represented by two quantities, the turbulence
intensity $q$ and the axial (streamwise) velocity $u$, sampled on the pipe
axis.  Specifically, $q$ is the magnitude of transverse fluid velocity (scaled
up by a factor of 6).  The centerline velocity $u$ is relative to the mean
velocity and is a proxy for the state of the mean shear that conveniently lies
between 0 and 1.  At low $\Rey$, as in Fig.~\ref{fig:DNS}(a), turbulence
occurs in localized patches propagating downstream with nearly constant shape
and speed. These are called equilibrium puffs~\citep{Wygnanski:1975,
Darbyshire:1995, Nishi:2008}, a misnomer since at low $\Rey$ puffs are only
metastable and eventually revert to laminar flow, i.e.\
decay~\citep{Faisst:2004p898, Peixinho:2006, Hof:2006p607, Willis:2007p833,
Schneider:2008p642, Hof:2008p914, Avila:2010p839,
Kuik:2010p911}. Asymptotically the flow will be laminar parabolic flow, $(q=0,
u=1)$, throughout the pipe.  For intermediate $\Rey$, as in
Fig.~\ref{fig:DNS}(b), puff splitting frequently occurs~\citep{Wygnanski:1975,
Nishi:2008, Moxey:2010, Avila:2011}.  New puffs are spontaneously generated
downstream from existing ones and the resulting pairs move downstream with
approximately fixed separation.  Further splittings will occur and
interactions will lead asymptotically to a highly intermittent mixture of
turbulent and laminar flow~\citep{Rotta:1956, Moxey:2010}.  At yet higher
$\Rey$, turbulence is no longer confined to localized patches, but spreads
aggressively in so-called slug flow~\citep{Wygnanski:1973, Nishi:2008,
Mullin:2011}, as illustrated in Fig.~\ref{fig:DNS}(c). The asymptotic state is
uniform, featureless turbulence throughout the pipe~\citep{Moxey:2010}.

\subsection{PDE model}

The modeling in \citet{barkley:2011} is based on the following physical
features of transitional turbulence in pipes.  At the upstream (left in
Fig.~\ref{fig:DNS}) edge of turbulent patches, laminar flow abruptly becomes
turbulent.  Energy from the laminar shear is rapidly converted into turbulent
motion and this results in a rapid change to the mean shear
profile~\citep{Wygnanski:1973, Hof:2010p65}.  In the case of puffs, the
turbulent profile is not able to sustain turbulence and thus there is a
reverse transition~\citep{Wygnanski:1973, Sreenivasan:1979} from turbulent to
laminar flow on the downstream side of a puff. In the case of slugs, the
turbulent shear profile can sustain turbulence indefinitely; there is no
reverse transition and slugs grow to arbitrary streamwise
length~\citep{Wygnanski:1973,Nishi:2008}.  On the downstream side of turbulent
patches the mean shear profile recovers slowly~\citep{Sreenivasan:1979}, seen
in the behavior of $u$ in Fig. 1.  The degree of recovery dictates
how susceptible the flow is to re-excitation into
turbulence~\citep{Hof:2010p65}.

The following partial-differential equation (PDE) model captures the essence
of these physical features:
\begin{eqnarray}
\partial_t q + U \partial_x q & = & q 
\left( u + r - 1 - (r + \delta) (q -1)^2  \right) + \partial_{xx} q, 
\label{eq:q_pde} \\
\partial_t u + U \partial_x u & = & 
\epsilon_1 (1 - u) - \epsilon_2 u q - \partial_x u. 
\label{eq:u_pde} 
\end{eqnarray}
In the model, the parameter $r$ plays the role of Reynolds number $\Rey$. $U$
accounts for downstream advection by the mean velocity, and is otherwise
dynamically irrelevant since it can be removed by a change of reference
frame. The model includes minimum derivatives, $q_{xx}$ and $u_x$, needed for
turbulent regions to excite adjacent laminar ones and for left-right symmetry
breaking.

The core of the model is seen in the $q$-$u$ phase plane in
Fig.~\ref{fig:pde}.  The trajectories are organized by the nullclines: curve
where $\dot u = 0$ and $\dot q=0$ for the local dynamics ($q_{xx} = q_{x} =
u_{x} = 0$).  For all $r$ the nullclines intersect in a stable, but excitable,
fixed point corresponding to laminar parabolic flow.  The $u$ dynamics with
$\epsilon_2 > \epsilon_1$ captures in the simplest way the behavior of the
mean shear. In the absence of turbulence ($q=0$), $u$ relaxes to $u=1$ at rate
$\epsilon_1$, while in response to turbulence ($q>0$), $u$ decreases at a
faster rate dominated by $\epsilon_2$.  Values $\epsilon_1 = 0.04$ and
$\epsilon_2 = 0.2$ give reasonable agreement with pipe flow.  The
$q$-nullcline consists of $q=0$ (turbulence is not spontaneously generated
from laminar flow) together with a parabolic curve whose nose varies with $r$,
while maintaining a fixed intersection with $q=0$ at $u=1+\delta$, ($\delta =
0.1$ is used here).  The upper branch is attractive, while the lower branch is
repelling and sets the nonlinear stability threshold for laminar flow. If
laminar flow is perturbed beyond the threshold (which decreases with $r$ like
$r^{-1}$), $q$ is nonlinearly amplified and $u$ decreases in response.

The (excitable) puff regime occurs for $r < r_c \simeq \epsilon_2/(\epsilon_1
+ \epsilon_2)$, Figs.~\ref{fig:pde}(a) and (c).  The upstream side of a puff
is a trigger front ~\citep{TYSON:1988p1143} where abrupt laminar to turbulent
transition takes place. However, turbulence cannot be maintained locally
following the drop in the mean shear. The system relaminarizes (reverse
transition) on the downstream side in a phase front~\citep{TYSON:1988p1143}
whose speed is set by the upstream front.  Following relaminarization, $u$
relaxes and laminar flow regains susceptibility to turbulent perturbations.
The slug regime occurs for $r > r_c$, Figs.~\ref{fig:pde}(b) and (d). The
nullclines intersect in additional fixed points.  The system is bistable and
turbulence can be maintained indefinitely in the presence of modified shear.
Both the upstream and downstream sides are trigger fronts, moving at different
speeds, giving rise to an expansion of turbulence.

\subsection{SPDE model}

While the PDE model captures the essence of the puff-slug transition, the
model of turbulence is too simple to capture features such as puff decay and
puff splitting. In \citet{barkley:2011}, a more realistic model was obtained
by employing a tent map to mimic shear turbulence.  The map was designed to
give a local phase-space structure similar to the nullcline picture for the
PDE seen in Fig.~\ref{fig:pde}, with the exception that the upper turbulent
branch is instead a region of transient chaos.  This approach was motivated by
the view that shear turbulence is locally a chaotic
saddle~\citep{Eckhardt:2007p887} and it naturally extends
previous ideas of modeling chaotic transients with maps
 \citep{Chate:1988p8,Bottin:1998p305,Vollmer:2009p1066}. 
The resulting model has the
advantage of being deterministic, as is fluid flow, at least at the level
of the Navier-Stokes equations.

Here I consider an alternative approach and model turbulence as noise. This is
at the other extreme from the low-dimensional map. Here the dynamics is
infinite dimensional and not deterministic.  The simplest approach is to
apply noise to the $q$ equation and assume it is proportional to $q$
itself. This leads to the following stochastic PDE (SPDE) model:
\begin{eqnarray}
\partial_t q + U \partial_x q & = & q 
\left( u + r - 1 - (r + \delta) (q -1)^2  \right) + \partial_{xx} q 
+ \sigma q \eta, \label{eq:q_noise} \\
\partial_t u + U \partial_x u & = & 
\epsilon_1 (1 - u) - \epsilon_2 u q - \partial_x u. 
\label{eq:u_noise} 
\end{eqnarray}
where $\eta=\eta(x,t)$ is Gaussian noise.  The parameter $\sigma$ controls the
noise strength.  In reality, shear turbulence has significant correlations on
the scale of a puff, but these correlations are not considered here and
$\eta(x,t)$ taken here to be space-time white.

A large advantage of modeling the effect of turbulence through a noise term is
that one has a direct connection to the simple PDE model.  Moreover, analysis
of the SPDE is likely to be easier than analysis of the deterministic map
model. The price is the loss of deterministic dynamics.

\begin{figure}[t]
\begin{minipage}{14pc}
\includegraphics[width=14pc]{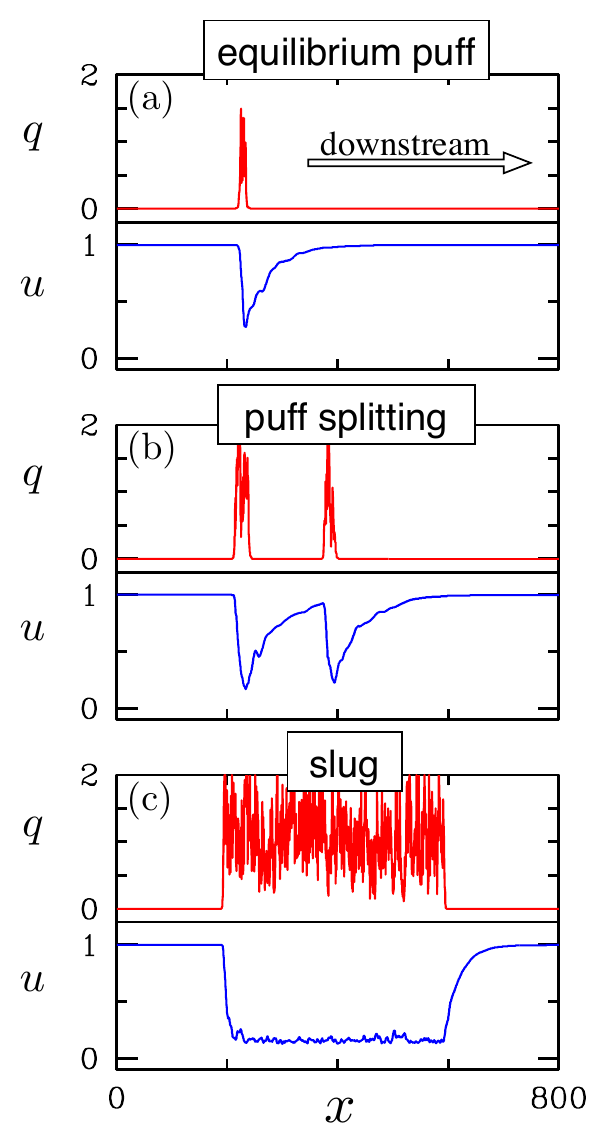}
\caption{\label{fig:model}
Three regimes of transitional pipe flow from simulations of the SPDE
~\eqref{eq:q_noise}-\eqref{eq:u_noise}.  Shown are instantaneous values of $q$
and $u$.  (a) Puff at $r = 0.7$.  (b) Puff splitting at $r = 0.94$.  (c) Slug
at $r = 1.2$.}
\end{minipage}\hspace{2.9pc}%
\begin{minipage}{21pc}
\includegraphics[width=21pc]{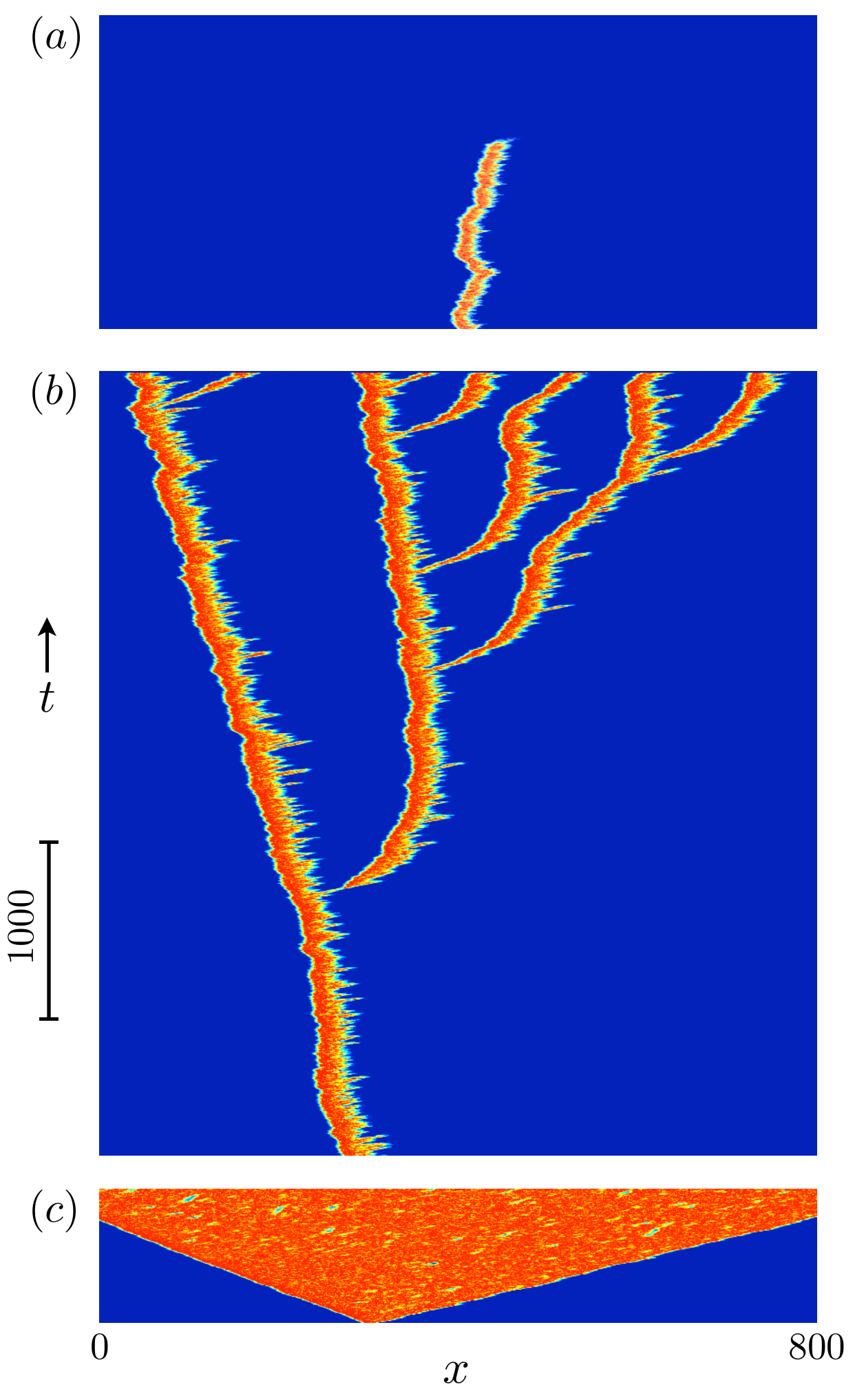}
\caption{\label{fig:spacetime}
Space-time diagrams illustrate (a) decaying puff at $r=0.7$, (b) puff
splitting at $r=0.94$, and (c) slug formation at $r=1.2$.  Turbulence
intensity $q$ is plotted on a logarithmic scale in a frame co-moving with
structures.  }
\end{minipage} 
\end{figure}

Figures~\ref{fig:model} and \ref{fig:spacetime} show the regimes of
transitional pipe flow from simulations of
Eqs.~\eqref{eq:q_noise}-\eqref{eq:u_noise}. The deterministic parameters are
as before: $\epsilon_1 = 0.04$, $\epsilon_2 = 0.2$, and $\delta = 0.1$. The
noise strength is $\sigma = 1.4$. Figure~\ref{fig:model} shows solution
snapshots in terms of the model variable $q$ and $u$. Puffs, puff splitting,
and slugs are found very similar to those observed in full DNS (see
Fig.~\ref{fig:DNS}) and in the deterministic map
model \citep[see][]{barkley:2011}. The dynamics of the different regimes is
seen in the space-time plots of Fig.~\ref{fig:spacetime}. At low $r$, puffs
are metastable. They persist for long times before abruptly decaying. For
intermediate $r$, puff
splitting occurs. New puffs are spontaneously nucleated downstream of existing
puffs and the system evolves to an intermittent mixture of turbulent and
laminar phases.
At larger $r$, slugs are observed which differ from the deterministic PDE
mainly in that they first occur at larger $r$ and the upper branch is noisy
rather than constant.
An investigation of the lifetime statistics of puff decay and puff splitting
in the SPDE is currently underway.

While all three regions shown in Figs.~\ref{fig:model} and \ref{fig:spacetime}
strongly resemble their counterparts in full DNS and experiment, the
splitting regime is particularly significant and worthy of further
comment. Unlike for puffs and slugs, which are essentially contained in the
model by construction, splitting is seen to arise naturally from the
elementary puff-slug transition in the presence of complex turbulent dynamics
(either noise as here or chaotic dynamics as in \citet{barkley:2011}).
The space-time plot in Fig.~\ref{fig:spacetime}(b) could easily be mistaken
for the corresponding plot from full DNS \citep[e.g. see][]{Avila:2011}.
Sufficient turbulence occasionally escapes from the irregular downstream side
of a puff to nucleate a new puff downstream. Visually, this is just as in
real pipe flow and is a strong qualitative validation of this modeling
approach.

\section{Model for plane Couette flow}

One of the main difficulties in extending these ideas to other shear flows,
such as channel flow or plane Couette flow, is the complexity of the mean flow
in these cases~\citep{Barkley:2007}.  It is not clear at the present time
whether one can adequately model the mean shear in these flows using simple
scalar fields.  Nevertheless, I discuss here some preliminary ideas on how
plane Couette flow might be approached within this modeling framework.

\begin{figure}[h]
\begin{minipage}{18pc}
\includegraphics[width=18pc]{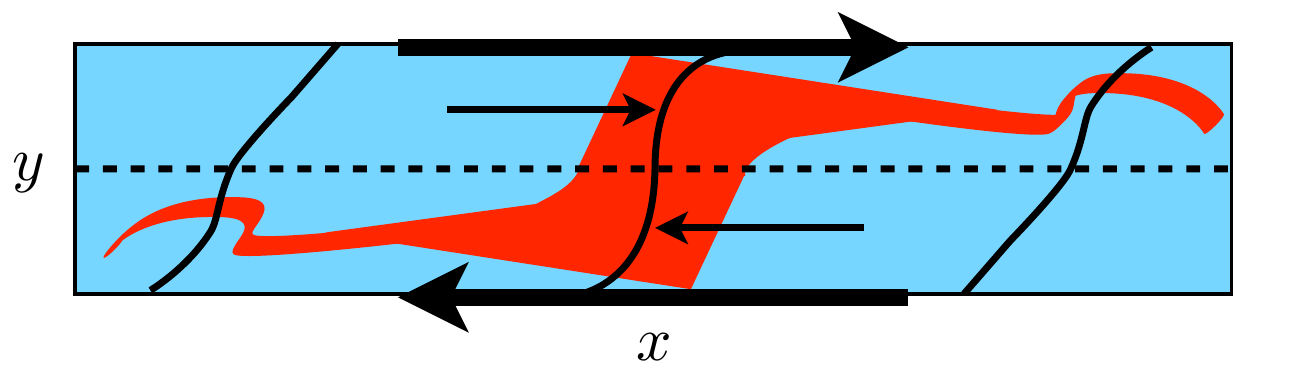}
\caption{\label{fig:sketch}
Sketch of plane Couette flow. The shaded region (red) represents a turbulent
patch surrounded by laminar flow. Arrows indicate the motion of the bounding
plates and the direction of the mean flow in the upper and lower halves of the
domain. Three shear profiles are sketched. The $x$-coordinate is centered on
the time-averaged flow which has centro-symmetry. }
\end{minipage}\hspace{2.9pc}%
\begin{minipage}{17pc}
\includegraphics[width=17pc]{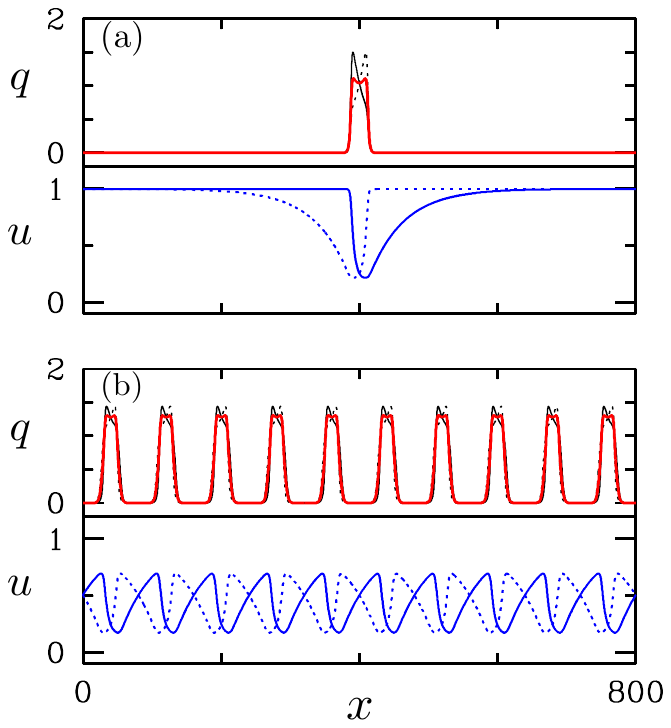}
\caption{\label{fig:planeCouette}
Solutions of the plane Couette model, Eqs.~\eqref{eq:q1}-\eqref{eq:u2}. (a) 
Localized state at $r=0.7$ and (b) periodic state at $r=0.9$.  Solid
(dotted) curves represent variables in the upper (lower) half of the
domain. The mean turbulence, $(q_1+q_2)/2$, is plotted in bold (red).  }
\end{minipage} 
\end{figure}

Figure~\ref{fig:sketch} shows a sketch for plane Couette flow. A turbulent
patch (red) is shown surrounded to the left and right by laminar flow. One can
view this as a cut through a single turbulent band in the striped regime or
through a localized patch of
turbulence~\citep{Prigent:2002,Barkley:2005p1065}. However, the model is
rather crude at present and the important three-dimensional aspects of the
problem are not taken into account.

Recall that the geometry of plane Couette flow has translation symmetry in the
streamwise direction $x$ and also centro-symmetry (rotation by $\pi$) about
any point on the midplane $y=0$. Taking the point to be at $x=0$, the
centro-symmetry is the transformation $(x,y) \to (-x,-y)$. Time averaged
turbulent-laminar patterns break translational symmetry, but do not break
centro-symmetry \citep{Barkley:2007}.

The idea is to model intermittent turbulence in plane Couette flow as two
layers, each described by one-dimensional equations similar to those for pipe
flow. The turbulence in the two layers is assumed to be coupled. Taking into
account that the mean advection in the bottom layer is opposite to that in the
top layer, I propose the following PDE model
\begin{eqnarray}
\partial_t q_1 + U \partial_x q_1 & = & q_1 
\left( u_1 + r - 1 - (r + \delta) (q_1 -1)^2  \right) + \partial_{xx} q_1 
+ \kappa (q_2 - q_1), \label{eq:q1} \\
\partial_t u_1 + U \partial_x u_1 & = & 
\epsilon_1 (1 - u_1) - \epsilon_2 u_1 q_1 - \partial_x u_1,  
\label{eq:u1} \\
\partial_t q_2 - U \partial_x q_2 & = & q_2 
\left( u_2 + r - 1 - (r + \delta) (q_2 -1)^2  \right) + \partial_{xx} q_2 
+ \kappa (q_1 - q_2), \label{eq:q2} \\
\partial_t u_2 - U \partial_x u_2 & = & 
\epsilon_2 (1 - u_2) - \epsilon_2 u_2 q_2 + \partial_x u_2.  
\label{eq:u2}
\end{eqnarray}
where $q_1$ and $u_1$ are the turbulence and mean shear in the upper layer and
$q_2$ and $u_2$ are the turbulence and mean shear in the lower layer.  These
equations are symmetric under translation in $x$ and the reflection defined by
$(x, q_1, u_1, q_2, u_2) \to (-x, q_2, u_2, q_1, u_1)$, which is the model
equivalent of centro-symmetry.

Figure~\ref{fig:planeCouette} shows solutions to
Eqs.~\eqref{eq:q1}-\eqref{eq:u2} with $U=1$ and the coupling constant $\kappa
= 0.1$. Other parameter values are the same as for the model pipe
simulations. For small $r$, stable localized states appear from localized
perturbations of laminar flow. At $r \simeq 0.75$ the localized states became
unstable and spread to form a periodic alternation of turbulent and laminar
phases. The localized and periodic structures both have centro-symmetry.

The choice of $U$ and the coupling parameter $\kappa$ are probably rather
important in obtaining steady patterns. The effect of noise is also not yet
fully understood as this investigation is still in a preliminary stage.

\section{Conclusion}

I have presented models of parallel shear flows in two scalar variables --
turbulence and mean shear. The model for pipe flow is based closely on
physical features of transitional turbulence and it reproduces nearly all
large-scale features of transitional pipe flow. The model for plane Couette
flow necessarily misses many features of the real flow since the model is only
one dimensional, whereas plane Couette flow has two extended
dimensions. Nevertheless, the plane Couette flow model is an important
starting point for further investigations of this and other parallel shear
flows such as plane channel and boundary layer flow.

%


\end{document}